\documentclass[10pt,conference]{IEEEtran}
\usepackage{graphicx}
\usepackage{amssymb}
\usepackage{epstopdf}
\usepackage{algorithmic}
\usepackage{hyperref}
\usepackage{verbatim}
\usepackage{paralist}
\usepackage{times}
\usepackage{color}
\usepackage{subfigure}
\usepackage[ruled,vlined,boxed]{algorithm2e}
\usepackage{balance}

\newcommand{\ignore}[1]{}

\newcommand{\mm}{{{\ensuremath\mu}MTU}}

\newcommand{\ContentFragment}{{\small \tt ContentFragment}}

\newcommand{\Name}{{\small \tt Name}}
\newcommand{\ContentObjectSize}{{\small \tt ContentObjectSize}}

\newcommand{\InternalState}{{\small \tt InternalState}}
\newcommand{\PayloadOffset}{{\small \tt PayloadOffset}}
\newcommand{\PayloadSize}{{\small \tt PayloadSize}}
\newcommand{\Payload}{{\small \tt Payload}}

\newcommand{\ContentDigest}{{\small \tt ContentDigest}}

\title{Secure Fragmentation for Content-Centric Networks (extended version)}
\author{\IEEEauthorblockN{Cesar Ghali}
\IEEEauthorblockA{University of California, Irvine\\
cghali@uci.edu}
\and
\IEEEauthorblockN{Ashok Narayanan}
\IEEEauthorblockA{Google\\
ashokn@ashokn.org}
\and
\IEEEauthorblockN{David Oran}
\IEEEauthorblockA{Cisco Systems\\
oran@cisco.com}
\and
\IEEEauthorblockN{Gene Tsudik}
\IEEEauthorblockA{University of California, Irvine\\
gts@ics.uci.edu}}

\date{}

\begin{document}
\maketitle
\thispagestyle{plain}
\pagestyle{plain}

\begin{abstract}
Content-Centric Networking (CCN) is a communication paradigm that emphasizes  
content distribution. Named-Data Networking (NDN) is an instantiation of CCN,  
a candidate Future Internet Architecture. NDN supports 
human-readable content naming and router-based content caching
which lends itself to efficient, secure and scalable
content distribution. Because of NDN's fundamental requirement that each content object must be 
signed by its producer, fragmentation has been considered incompatible
with NDN since it precludes authentication of individual content fragments by routers. 
The alternative is to perform hop-by-hop reassembly, which incurs prohibitive delays.
In this paper, we show that secure and efficient content fragmentation is both possible and even
advantageous in NDN and similar content-centric network
architectures that involve signed content. We design a concrete
technique that facilitates efficient and secure content fragmentation in NDN, discuss its
security guarantees and assess performance. We also describe a prototype implementation
and compare performance of cut-through with hop-by-hop fragmentation and reassembly.
\end{abstract}

\section{Introduction}\label{intro}
The Internet is a {\em de facto} public utility used by a significant
fraction of the humankind who rely on it for numerous daily activities.
However, despite unparalleled success and unexpected longevity, 
the current TCP/IP-based protocol architecture may be obsolete.
To this end, several research efforts to design a next-generation Internet 
architecture have sprung up in recent years \cite{NSF-FIA}.

One key motivator for a new Internet architecture is the fundamental
shift in the nature of traffic: from the mainly low-bandwidth 
interactive (e.g., remote log-in) and store-and-forward (e.g., email) nature of 
the early Internet to the web-dominated Internet of today. At the same time,
massive and rapidly-increasing amounts of content are 
produced and consumed (distributed) over the Internet. This transpires over
social networks such as Facebook, media-sharing sites such as YouTube and
productivity services such as GoogleDocs. Consequently, the emphasis of
Internet communication has shifted from telephony-like conversations between two 
IP interfaces to a consumer who wants content delivered rapidly and securely, regardless of 
where it comes from.  This motivates reconsidering the Internet architecture.

Content-Centric Networking (CCN) \cite{gritter2001architecture, Jacobson2009, koponen2007data} 
is an approach to (inter-)networking designed for efficient, secure and scalable 
content distribution \cite{Jacobson2009}.
In CCN, named content -- rather than named interfaces or hosts -- is treated as a first-class 
entity. Named-Data Networking (NDN) \cite{NDN} is an example of CCN that 
stipulates that each piece of named content must be signed by its producer (also known as publisher). 
This allows decoupling of trust in content from trust in an entity (host or router) that might store and/or 
disseminate that content. These NDN features facilitate in-network caching of content to 
optimize bandwidth use, reduce latency, and enable effective simultaneous utilization of multiple 
network interfaces.

NDN is an on-going research effort and one of several architectures being considered as a 
candidate future Internet architecture. Other such efforts include: 
ChoiceNet \cite{choicenet}, XIA \cite{xia}, Mobility-First \cite{mobilityfirst} and Nebula \cite{nebula}. 
Regardless of which, if any, approach 
eventually succeeds, all of them need to
address some of the same issues, such as naming/addressing, routing/forwarding and security/privacy. 

\subsection{Fragmentation:}
One issue that straddles both networking (specifically, packet forwarding) and security 
is fragmentation of large packets. Originally present in IPv4 \cite{ipv4}, intermediate fragmentation was 
deprecated in IPv6 \cite{ipv6}, for a number of reasons, many of which were identified in
by Mogul \cite{mogul}, e.g., router overhead and code complexity. Also, there have been 
attacks that took advantage of IPv4 reassembly \cite{ziemba1995security}. Thus, eschewing 
fragmentation made sense for IPv6. However, the same 
might not hold for all network architectures. We show in Section~\ref{ndnfrag} that, for some very different 
types of networking, such as CCN/NDN, fragmentation is sometimes unavoidable and might 
even be beneficial.

\subsection{Focus:}
This work represents the first exploration of efficient and secure fragmentation in the context of CCN/NDN.
Its primary value is the construction of a secure fragmentation scheme which
addresses several important security and efficiency issues. (Section~\ref{related-fragmentation-icn} 
describes current handling of fragmentation in CCN/NDN.)

The intended contribution of this paper is two-fold: 
{\bf First,} we discuss, in detail, numerous issues related to fragmentation of both 
interest and content packets in NDN and arrive at the following key conclusions:
\begin{compactitem}
\item Interest fragmentation is unavoidable: if encountered, hop-by-hop reassembly is a must.
\item Content fragmentation is similarly unavoidable.
\item Minimal MTU discovery helps, but does not obviate the need for fragmentation.
\item Intermediate reassembly is viable but buffering can be costly and latency is problematic.
\item Intermediate re-fragmentation is also unavoidable for content fetched from router caches. 
\item Reconciling cut-through forwarding of fragments (no intermediate reassembly) with content 
authentication is possible in an efficient manner.
\end{compactitem}
{\bf Second,} we construct a secure fragmentation and reassembly method for 
content-centric networks architectures, exemplified by NDN. 
We call this method {\em Fragmentation with Integrity Guarantees and Optional
Authentication} or FIGOA. It supports fragmentation of content packets at NDN
network layer and allows cut-through switching in routers by avoiding
hop-by-hop fragmentation and reassembly, thus lowering end-to-end delay.
However, even though cut-through switching reduces latency, it allows 
fragments to be temporarily stored in routers awaiting verification by FIGOA. This 
might result in exhausting routers resources if fragment-drop rate increases. To 
solve this issue, routers adaptively set time-outs for temporarily stored fragments. 
(This issue is not discussed further, as it is outside the scope of this paper.)

FIGOA is fundamentally compatible with the NDN's tenet of not securing the channel but
rather the content flowing through it. As its basis, FIGOA employs a delayed authentication 
method similar to \cite{Tsudik89}, which allows routers to verify signed content based on
out-of-order arriving fragments. Furthermore, FIGOA supports nested 
(successive or recursive) fragmentation. In the event that a given content ultimately fails
either integrity or authenticity check, reassembly and eventual delivery of corrupt content to 
consumers is prevented.

\paragraph{Organization} 
Section~\ref{overview} provides some background on NDN.
Then, Section~\ref{frag} summarizes fragmentation in IP. 
Fragmentation issues in NDN are discussed Section~\ref{ndnfrag},
The proposed FIGOA scheme is presented in Section~\ref{securefrag}.
Section~\ref{implementation} describes its prototype implementation 
which is then evaluated in Section \ref{evaluation}. Next,
Section~\ref{related} overviews related work. The paper ends with
the summary and future work in Section~\ref{end}.

\section{NDN Overview}\label{overview}
This section overviews NDN. In case of familiarity with NDN, it can be skipped with no lack of continuity.

NDN supports two types of packets: {\em interest} and {\em content}~\cite{ccnx-protocol}. The latter contains 
a human-readable name, actual data (content), and a digital signature over the packet computed by the content producer.
Names are hierarchically structured, e.g. \verb|/ndn/usa/cnn/frontpage/news| 
where ``\verb|/|'' is the boundary between name components. An interest packet contains the name of 
the content being requested, or a name prefix, e.g. \verb|/ndn/usa/cnn/| is a prefix of 
\verb|/ndn/usa/cnn/frontpage/news|. In case of multiple contents under a given name prefix, 
optional control information can be carried within the interest to restrict the content returned. Content
signatures provide data origin authenticity, however, trust management between a key
and a name prefix is the responsibility of the application.

All NDN communication is initiated by a (content) consumer that sends  
an interest for a specific content. NDN routers forward this interest towards the content producer 
responsible for the requested name, using name prefixes (instead of today's IP prefixes) for
routing. A {\em Forwarding Information Base} (FIB) is a lookup table used to determine interfaces 
for forwarding incoming interests, and contains [{\em name\_prefix}, {\em interface}] entries. 
Multiple entries with  the same {\em name\_prefix} are allowed, supporting multiple paths 
over which a given {\em name\_prefix} namespace is reachable. Akin to an IP forwarding table, 
FIB can be populated either by a routing protocol or manually.

NDN communication adheres to the ``pull'' model, whereby content is delivered to consumers only 
following an explicit request (interest). NDN content includes several fields. In this paper, we are 
only interested in the following:
\begin{compactitem}
\item \texttt{Signature} -- a public key signature, generated by the content 
producer, covering the entire content, including all explicit components of 
the name and a reference to the public key needed to verify it.
\item \texttt{Name} -- a sequence of explicit name components followed by an 
implicit digest (hash) component of the content that is recomputed at every 
hop. This effectively provides each content with a unique name and guarantees 
a match with a name provided in an interest. However, in most cases, the hash 
component is not present in interest packets, since NDN does not provide any 
secure mechanism to learn a content hash a priori.
\item \texttt{KeyLocator} -- a reference to the public key required to verify 
the signature. This field has three options: (1) verification key, (2) certificate 
containing the verification key, or (3) NDN name referencing the content that 
contains the verification key.
\end{compactitem}
Each NDN router maintains a Pending Interest Table (PIT) -- a lookup table containing 
outstanding [{\em interest}, {\em arrival-interface(s)}] entries. The first component of each
entry reflects the name of requested content, and the second -- a set
of interfaces via which interests for this content have arrived. 
When an NDN router receives an interest, it looks up its PIT to determine
whether an interest for the eponymous content is pending. There
are three possible outcomes: 
\begin{compactenum}
\item If the same name is already in the router's 
PIT and the arrival interface of the present interest is already in the set of 
{\em arrival-interfaces}  of the corresponding PIT entry, the interest is discarded. 
\item  If a PIT entry for the same name exists, yet the arrival interface is new, the
router updates the PIT entry by adding a new interface to the set; the interest is not forwarded further.
\item Otherwise, the router looks up its cache (called Content Store) for a matching content. 
If it succeeds, the cached content is returned and no new PIT entry is needed. Conversely, if 
no matching content is found, the router creates a new PIT entry and forwards the interest using its FIB.
\end{compactenum}
An interest might reach the actual content producer if no corresponding content has been
cached by any intervening router on the path between consumer and producer. Upon receipt of the 
interest, the producer 
injects requested content into the network, thus {\em satisfying} the interest. The content is then 
forwarded towards the consumer, traversing -- in reverse -- the path of the preceding interest. Each router 
on the path flushes state (deletes the PIT entry) containing the satisfied interest and 
forwards the content out on all arrival interfaces of the associated interest.
In addition, each router may cache a copy of forwarded content. Unlike their IP counterparts, NDN routers can 
forward interests out on multiple interfaces in order to increase likelihood
of fastest content retrieval. 

Not all interests result in content being returned. If an interest encounters either: (1) a router that 
can not forward it further or (2) a content producer that has no such content, no error packets are
generated. PIT entries in intervening routers simply expire if content can not be retrieved.
The consumer (who also maintains its local PIT) can choose to re-issue the same
interest after a timeout.

\section{Fragmentation Synopsis}\label{frag}
We define {\em fragmentation} as a means of splitting large packets into smaller packets, at the
network layer, independent of content publisher and without changing
any actual {\em content}. This is in contrast with {\em segmentation} where a content publisher 
splits a large content object into smaller ones, signing and naming each separately.
TCP/IP has an analogous dichotomy: TCP {\em segments} a byte stream into IP packets, whereas, 
IPv4 can {\em fragment} IP packets into smaller packets in order to fit them into a link MTU.

Since late 1980s, network-layer fragmentation has been widely
considered to be a headache and something to be avoided, based
primarily on the IPv4 experience \cite{mogul}. To understand this further, we
briefly discuss IPv4 fragmentation concepts.

Packet fragmentation is not a singular concept; it can be
divided into two types: source-based and network-based. Source-based
fragmentation is performed exclusively by the sender and is relatively
simple: assuming knowledge of the Maximum Transmission Unit (MTU) for
a given path to the destination, the source can fragment a packet with
almost\footnote{Dynamic routing in IP may cause successive packets
to take different paths, affecting the source's perceived MTU.} no fear that further
fragmentation might be encountered along the path. Of course,
knowledge of the MTU does not come for free; an MTU discovery protocol
is needed, e.g., \cite{RFC4821}. Also, the whole premise of
source-based fragmentation is questionable: 
Why should the source  fragment a large IP packet instead 
of simply ``segmenting'' it into a sequence of separate IP packets \cite{RFC2923}?  
Source-based segmentation often allows for more efficient use of smaller
datagrams; for example, segmented TCP datagrams can be individually
ACKed, whereas a larger TCP segment split using IP fragmentation can
only be processed as a whole by TCP. 

Network-based fragmentation requires routers to support extra
functionality (i.e., additional code) which entails appreciable processing
overhead \cite{mogul}. Having to fragment a packet takes a router off its
critical path and can thus cause congestion; this can also be exploited as
denial-of-service attack. Nonetheless, at a conceptual
level, it can be claimed that intermediate fragmentation offers better
bandwidth utilization than its source based counterpart, or no
fragmentation at all. 

Further issues are prompted by reassembly of fragments. In IPv4,
reassembly takes place only at the destination. Each IP packet is
allocated a buffer that stores fragments that have arrived thus far;
possibly, out-of-order. Once all fragments are received, the packet is
physically reassembled and passed on to the upper layer.  This
seemingly simple procedure has been a source of many attacks and
exploits \cite{ziemba1995security}. Reassembly by intermediate
hops/routers is not viable in IP, since fragments of the same packet
are not guaranteed to travel over the same path.

\section{Fragmentation in NDN}\label{ndnfrag}
NDN architecture does not provide explicit support for in-network fragmentation~\cite{NDNreport2013}. 
However, current NDN implementation, that runs as part of the NDN testbed~\cite{ndntestbed}, 
is implemented as an overlay on top of TCP or UDP. In this a setup, fragmentation is 
handled by either: (1) transport layer protocols, i.e.,TCP segmentation, or by (2) network layer 
protocols, i.e., IP fragmentation. Moreover, if NDN is running directly over the link layer, 
protocols such as NDNLP~\cite{NDNLP} can be used to handle fragmentation.  (See Section
~\ref{related-fragmentation-icn} for details.) The main drawback of these methods is 
that they all need reassembly at every hop.

The rest of this section discusses certain factors motivating fragmentation in NDN.
Using terminology in Figure \ref{tab:notation-orig}, fragmentation is considered in the context
of interest and content packets, respectively.

\begin{figure}[ht]
\fbox{\begin{minipage}{0.975\columnwidth}
\small
\begin{compactitem}
\item[\bf Maximum Transmission Unit (MTU):]  largest unit (packet) size for 
network-layer transmission over a given link between two adjacent nodes.
\item[\bf Publisher or Producer:] an entity that produces and signs content;
we use these two terms interchangeably.
\item[\bf Consumer:] an entity that requests (consumes) content.
\item[\bf Router:] a network-layer entity that routes content but neither 
produces nor consumes it (except perhaps for routing information).
\item[\bf Content object (CO):] a unit of NDN data, named and signed by its producer/publisher.
\item[\bf Content fragment (CF):] a unit of NDN network layer transmission; content fragment is 
the same as content object if the latter fits within the MTU of a link between two adjacent routers. 
\item[\bf Segmentation:] a process of partitioning large content into separate content objects by explicitly
naming and signing each one. Can be performed only by a producer of content.
\item[\bf Fragmentation:] a process of splitting an (already signed and named) content object 
into multiple content fragments. Can be performed by a producer, a router or 
any other NDN entity that produces, stores or caches content.
\item[\bf Re-fragmentation:] a process of splitting a fragment of a content object into multiple
fragments. Sometimes referred to as inter-network fragmentation \cite{mogul}.
Can be performed by a router. 
\item[\bf Reassembly:] a process of re-composing a content object from its fragments.
Can be performed by a consumer or a router (in case of intermediate reassembly).
\item[\bf Fragment Buffering:] a process of maintaining a stash of fragments until complete 
packet reassembly becomes possible.
\item[\bf Cut-Through Switching (of fragments):] forwarding of individual content 
fragments without reassembly.
\end{compactitem}
\end{minipage}}
\caption{Terminology}
\label{tab:notation-orig}
\end{figure}
\normalsize

\subsection{Fragmentation of Interests}
As discussed in Section~\ref{overview}, an interest packet carries
the name of content requested by the consumer. NDN does not 
mandate any confidentiality, integrity or authenticity requirements for 
interest packets. 
Due to no limitation on the length of content names, it is
quite plausible that an interest packet carrying a very long name might 
not fit into a network-layer MTU, thus prompting the need for source-based
and/or intermediate fragmentation. Fortunately, this does not pose any real challenges, since
the ``design space'' of interest fragmentation is very confined. Specifically, we claim that:

\medskip \centerline{\fbox{ \begin{minipage}{0.45\textwidth}
\it
If interest packets are fragmented and, possibly re-fragmented,
hop-by-hop (intermediate) reassembly of fragmented interest packets is a necessity.
\end{minipage}}} \medskip

This claim is easy to support because, as described in Section~\ref{overview},
each router that receives an interest must look up its PIT and/or cache using
the name carried in that interest. If the name itself spans multiple fragments 
which are processed independently (without reassembly), such lookups  
are infeasible.

Furthermore, since the consumer issuing an interest has no {\em a priori} knowledge
of the smallest MTU on the path to the closest copy of requested content,
it can not pre-fragment an interest in order to avoid further fragmentation by
intermediate routers, unless there is a well-defined global minimum MTU for
NDN interests.

Based on the above, for the remainder of this paper, we assume source-based
(and possibly intermediate) fragmentation coupled with intermediate reassembly 
for interests. The remaining discussion of fragmentation is limited to NDN 
content packets. 

\subsection{Fragmentation of Content}\label{fragmentation-of-content}
Recall that NDN mandates each content to be signed by its producer.
This means that, in principle, any NDN entity, whether router or consumer\footnote{
Content signature verification is mandatory for consumers and optional for routers.}
is able to check content integrity and authenticity, based on the producer's
public key, itself embedded in a separate signed content (a de facto credential or certificate).
The public key can be either referred to by name in the content header,  or
enclosed in its entirety as part of the content. 

Consequently, in order to abide by NDN tenets, fragmentation must not preclude routers
from verifying signatures, i.e., checking authenticity and integrity of content.
This speaks in favor of either: (1) no intermediate fragmentation at all, or (2) intermediate 
(hop-by-hop) reassembly.

\subsubsection{\bf Producer-based Fragmentation or Segmentation}
At the first glance, there seems to be no reason for a content producer to fragment
large content. Instead, it can simply {\em segment} it into individually named and 
separately signed content objects. This segmentation approach is sensible for content meant
to be pushed (e.g., email) or generated dynamically, e.g., in response to  
a database query. Segment size can be determined from an MTU discovery 
protocol (see Section~\ref{interfrag} below). This would assure no intermediate
fragmentation.

However, for content that is meant to be pulled (distributed),
a producer may benefit from signing and naming it once and not worrying about
repeating (possibly expensive) segmentation procedure each time it receives an interest 
for the same content. In this case, when an interest arrives, the producer may
choose to fragment a previously produced content object.
This entails no real-time cryptographic overhead. Alternatively, a producer could
choose to segment content using the smallest MTU of all of its interfaces,
thus incurring even less processing at interest arrival time.

An important issue is the content header overhead incurred when generating 
small-size segments. 
Segmenting a large object into many MTU-sized segments requires each of them to have 
its own header, dominated by the Signature component which contains a number of 
fields. 

\begin{figure}[t]
\centering
\includegraphics[width=0.7\columnwidth]{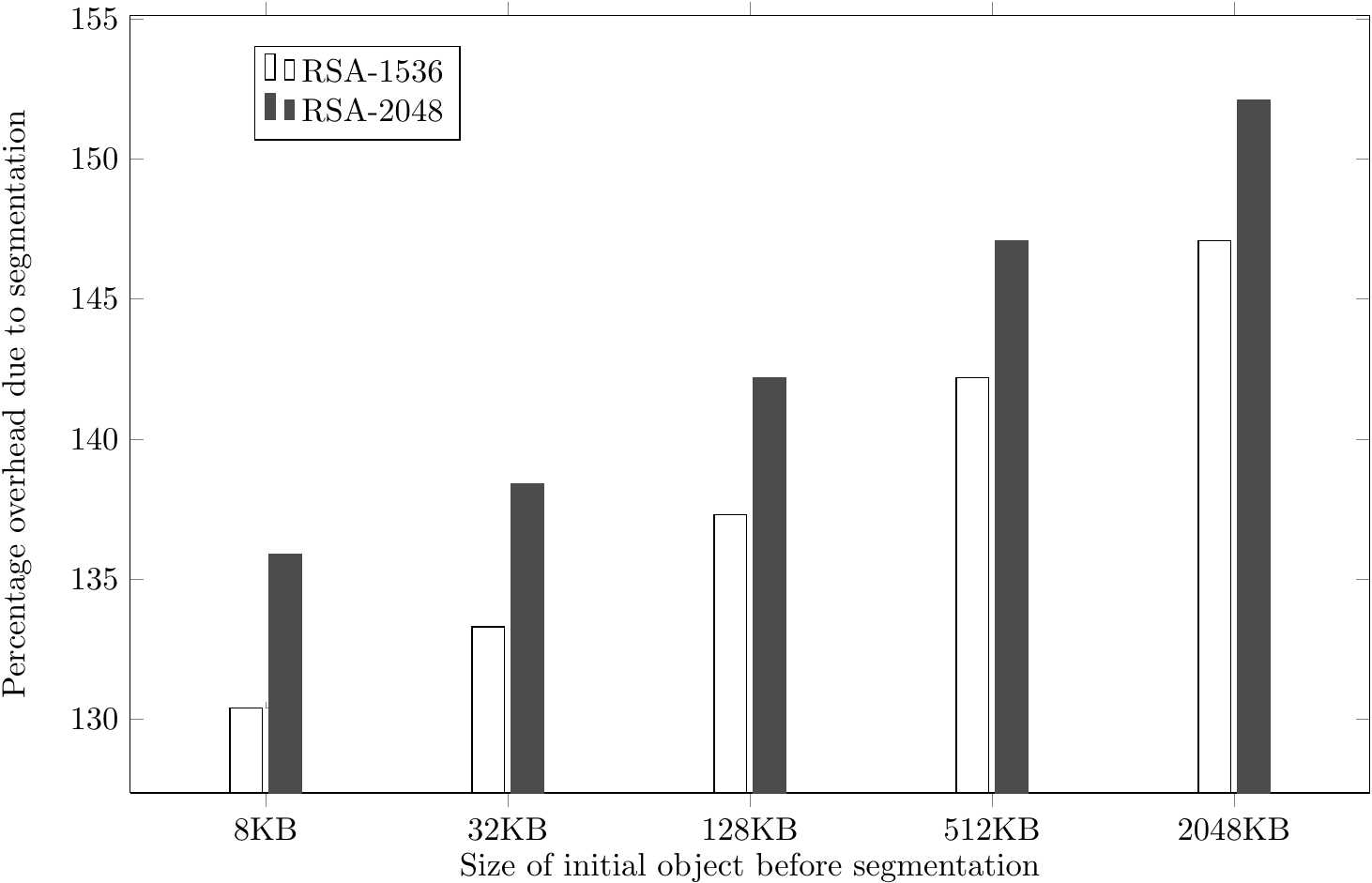}
\caption{Byte count overhead for small signed segments}
\label{fig:signovh}
\end{figure}

Without getting into details of NDN signature format, Figure~\ref{fig:signovh} 
shows the overhead of segmenting larger objects down to MTU size. We use a standard 
$1,500$-byte link MTU and SHA-256 as the hash algorithm.
We considered both RSA-1536 and RSA-2048 signatures. The Signature field therefore contains:
12 bytes of fixed overhead (headers), actual signature bits (192 bytes for RSA-1536,
256 bytes for RSA-2048).
However, estimating the exact size of the signature field is
more complex. This is because the \texttt{KeyLocator} field 
(which is part of the signature field) can be of arbitrary size (and if
it carries a certificate, it can be {\em very} large). For now,
we assume a small 20-byte \texttt{KeyLocator}, along with an SHA-256
hash. Figure~\ref{fig:signovh} shows that there is a definite penalty for segmenting at
the publisher. Even in the most favorable case (8KB data objects,
RSA-1536), over 30\% of the bits are wasted on redundant
information. As we move to larger objects,
this overhead can grow to 50\%.

\subsubsection{\bf Whither Intermediate Fragmentation\label{interfrag}}
Regardless of whether a producer segments or fragments content, intermediate 
fragmentation can not avoided or ruled out, since NDN does not mandate a globally minimal MTU.
Even if it existed, segmenting content to adhere to this MTU might be very wasteful 
due to poor bandwidth utilization (on links that have higher MTUs) and
cryptographic overhead due to increased costs of signature generation by producers and
signature verification by consumers as well as (optionally) by routers. 

Another possibility is to introduce an MTU discovery method, whereby, for example,
an interest traveling towards requested content could have a new field reflecting the smallest
MTU (\mm) discovered thus far on its path.\footnote{This is actually the MTU of the {\em opposite} 
link direction from the direction the interest traveled - links may have asymmetric MTUs.} 
This is a viable and light-weight approach, particularly because, in NDN, 
a content must traverse, in reverse, the very same path taken by an interest 
for that content. Hence, when the first entity that stores, caches or produces 
requested content receives an interest, it can use \mm\ to fragment 
(or segment, if this entity is the producer). Note that ``entity'' could encompass: 
(1) an application-level repository that stores content it does not produce, (2) a 
router that caches content or a (3) producer/publisher that generates its own content. 
This way, fragmentation would occur only once per interest. 

However, fragmentation via interest-based \mm\ discovery does not eliminate the need for 
re-fragmentation. Consider the following scenario:
\begin{compactenum}
\item Consumer $A$ issues interest $int_{A}$ for $CO$ to router $R$.
\item $R$ receives and marks $int_{A}$ with $\mm = MTU_{(R \rightarrow A)}$ (MTU corresponding $R-A$ link). 
It then creates a PIT entry for $int_{A}$.
\item $R$ forwards $int_{A}$ to adjacent producer $P$.
\item Since $MTU_{(P \rightarrow R)} > \mm_{(R \rightarrow A)}$, $P$ does not change $\mm$ in $int_{A}$.
\item $P$ immediately satisfies $int_{A}$, fragmenting $CO$ according to $\mm$.
\item Meanwhile, between Step 3 and now, consumer $B$ issues interest $int_{B}$ and forwards it to $R$.
\item $R$ receives $int_{B}$ and marks it with $MTU_{(R \rightarrow B)}$ where $MTU_{(R \rightarrow B)} < \mm$. 
$R$ collapses $int_{B}$ into existing PIT entry for $int_{A}$. At this time $R$ is buffering 
fragments which have arrived from $P$, however, not all fragments of $CO$ have arrived yet.
\item $R$ partially satisfies $int_{B}$ using fragments available in the buffer, previously forwarded to $A$. These 
fragments are re-fragmented with $MTU_{(R \rightarrow B)}$. Any further fragments which arrive from $P$ 
are also re-fragmented by $R$ to $B$ using $MTU_{(R \rightarrow B)}$.
\end{compactenum}
Despite the fact that \mm\ discovery does not eliminate re-fragmentation, it  is practically 
free in terms of extra processing and bandwidth overhead. More importantly, it results in 
less re-fragmentation, since it assures that re-fragmentation occurs {\bf at most once} for 
each collapsed interest at each intermediate router. This can be particularly advantageous 
in the case of monotonically shrinking MTUs where re-fragmentation must occur at each 
hop. With \mm, this is curtailed at the source of content (which is either some intermediate 
router or the producer) due to pre-fragmentation.

\subsection{Considering Intermediate Reassembly} \label{reass}
We now discuss intermediate reassembly. There are at least two factors that motivate it. 

First, we consider the case of increasing MTUs on links that compose the reverse
path taken by content fragments on the way to the consumer. If MTUs increase 
monotonically, it might make sense to reassemble fragments (at least partially) 
to obtain better bandwidth utilization. However, this benefit is arguably  
outweighed by costs incurred by reassembly, i.e., processing, memory and code in 
routers. (Recall that, as in today's IPv4, reassembly would require a router 
to maintain a separate buffer pool for all fragments of a given content received so far.)

The second factor is security: if a fragment does not carry the content producer's 
signature, how can a router check its authenticity? As mentioned earlier, NDN stipulates
that routers, though not required to do so, must be able to verify content signatures.
As we argued above, it seems infeasible for the producer to pre-fragment
or pre-segment content such that each possible future fragment of that content
would carry the producer's signature. 

Hop-by-hop reassembly of content fragments would clearly solve the problem and
address both factors mentioned above. With it, a router would receive fragments in arbitrary 
order and neither cache nor forward them until all fragments arrive. It would 
then reassembles them and verify the content signature with the producer's public key. 
(See Section~\ref{authentication} for more details.)

The main problem with hop-by-hop reassembly is increased end-to-end
latency, resulting into lower throughput for adaptive algorithms, such as TCP.
If multiple flows are passing through the router, the fairest distribution of latency overhead is to
interleave fragments, as in MLPPP LFI \cite{rfc1990}. This interleaving causes 
significant latency between consecutive fragments of an object, which grows with the number of
simultaneous flows. Latency accumulates at each hop, since
all fragments need to be reassembled and then re-fragmented for
transmission. The alternative is cut-through fragment forwarding,
where each fragment is forwarded immediately, as it arrives.

We attempt to evaluate the benefits of cut-through fragment forwarding
by considering a simple topology with a linear $8$-hop path with $100$ Mb/s
links. Each link accumulates $10$ms of latency, ignoring intra-hop and
queuing delays for now. We assume $8,400$-byte content objects 
split into $7$ fragments of $1,300$ bytes each.

\begin{table}[t]
\centering
\small
\resizebox{0.5\textwidth}{!}{
\begin{tabular}{|c|c|c|c|c|c|c|}
\cline{2-7} 
\multicolumn{1}{c}{} & \multicolumn{6}{|c|}{\textbf{Number of flows}} \\ \cline{2-7}
\multicolumn{1}{c|}{} & \bf 5 & \bf 10 & \bf 20 & \bf 30 & \bf 50 & \bf 100 \\ 
\hline\cline{1-7}
Inter-fragment gap (ms) & 0.52 & 1.04 & 2.08 & 3.12 & 5.20
& 10.4 \\
\hline
First-to-last fragment gap  (ms) & 3.22 & 6.34 & 12.58 &
18.82 & 31.30 & 62.50 \\
\hline
E2E latency: reassembly (ms) & 105.79 & 130.75 &
180.67 & 230.59 & 330.43 & 580.03 \\
\hline
E2E latency: cut-through (ms) & 83.22 & 86.34 & 92.58
& 98.82 & 111.30 & 142.50 \\
\hline
Reassembly slowdown \%-age & 127.12 & 151.43 & 195.14
                                & 233.34 & 296.87 & 407.03 \\
\hline
\end{tabular}
} 
\caption{Latency due to per-hop content reassembly}
\label{tab:reassem}
\end{table}

\begin{figure}
\centering
\includegraphics[width=0.8\linewidth]{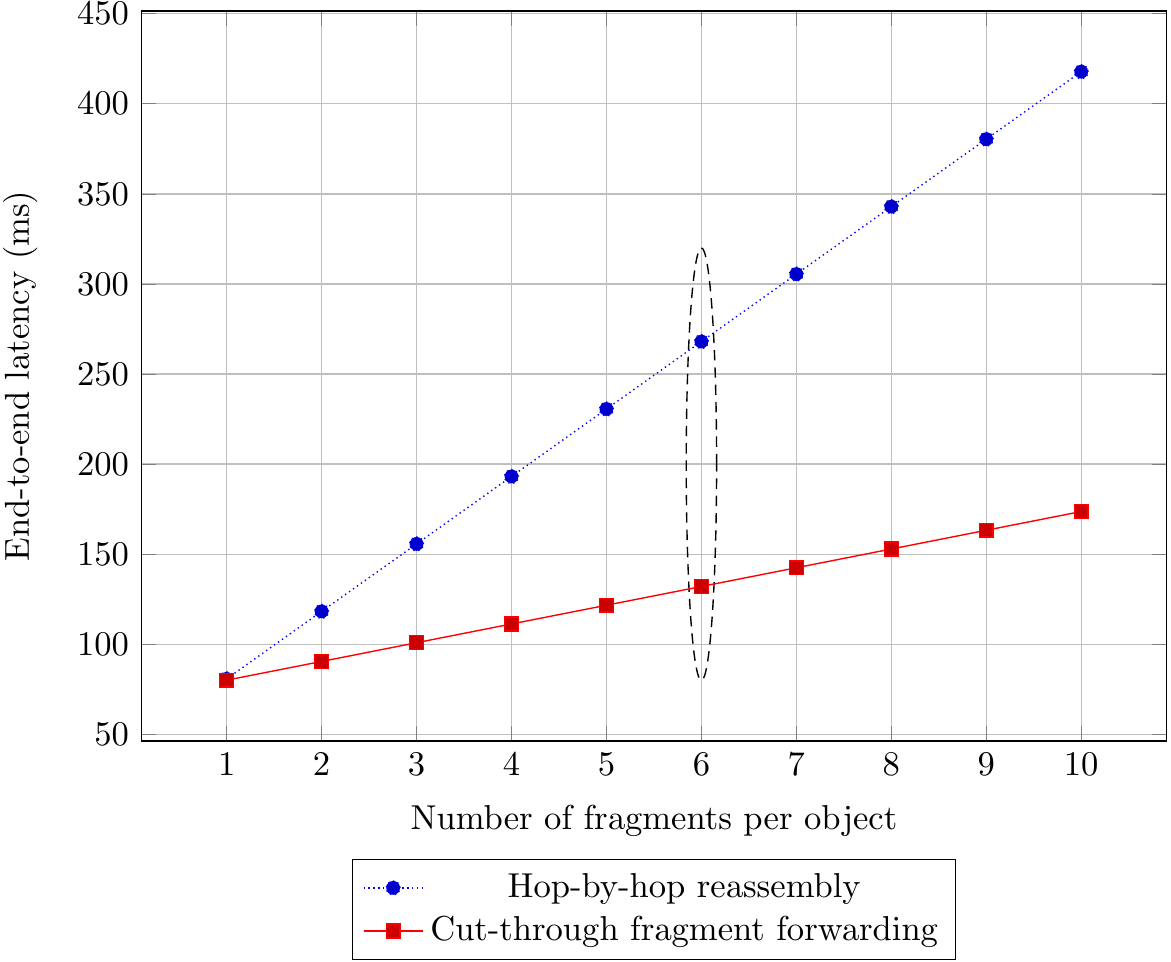}
\caption{Latency with different fragment counts per object}
\label{fig:latencyfrag}
\end{figure}

Table~\ref{tab:reassem} shows the slowdown caused by intermediate reassembly
as each node waits for all fragments of an object, for varying numbers of parallel 
flows (which controls the amount of interleaving). {\em Inter-fragment gap} is the time elapsed
between consecutive fragments of an object, caused by fragments of
other objects being interleaved. {\em First-to-last fragment gap} is the time elapsed between the
arrival of the beginning of the first fragment and the end of the last fragment. 
{\em E2E latency -- reassembly} is the total latency for each content object, with intermediate reassembly. 
{\em E2E latency --  cut-through} is the total latency in case of an object fragmented at the first hop and 
all fragments cut-through forwarded with no re-fragmentation or
reassembly in route. Finally, {\em Reassembly Slowdown} shows the
extra cost of reassembling and refragmenting at every hop, as
compared with cut-through forwarding. 

Figure~\ref{fig:latencyfrag} shows the evolution of increased latency
for various object sizes and fragment counts. We re-use the 8-hop 
topology described above and vary the number of flows on each link. 
Two links (close to the ends) have 10 flows across them, two links have 20 flows, 
two links have 50 flows and the two core links have 100 flows. The graph
shows that even a small number of fragments can significantly increase
latency over commonly seen path lengths and flow counts. It takes only
6 fragments per object to {\bf double} end-to-end latency of
hop-by-hop reassembly when compared with cut-through forwarding of fragments.
This clearly shows that any fragmentation scheme that
requires hop-by-hop reassembly of every content object (as is the case today
with CCNx \cite{ccnx} and NDNLP~\cite{NDNLP}) incurs 
severe penalties in a wide-scale deployment. We
believe that content object fragments must be forwarded in a cut-through
fashion; consequently, our scheme implements this feature.
 
At this point, it is worth asking: should routers perform reassembly and verify signatures? And if yes, which ones?
We believe that it does not make much sense for backbone routers to do so since potential attacks
are not likely to originate in the backbone, but rather at the edges of the Internet.
Whereas, signature verification at stub AS (ingress) routers is more appropriate, 
e.g., because of policy dictating that no fraudulent content must reach consumers.\footnote{Even 
though NDN stipulates consumer-based signature verification.} Also, stub AS egress routers might
reassemble fragments and verify signatures if there is a policy disallowing any fraudulent
content to exit an AS, e.g., for reasons of liability.

The above discussion yields a trivial observation that reassembly implies 
ability to verify signatures. However, it is unclear whether signature verification
implies the need for reassembly. This triggers the following challenge 
which we attempt to address in the remainder of this paper:

\medskip \centerline{\fbox{ \begin{minipage}{0.45\textwidth}
\it
If content objects are fragmented (and, possibly re-fragmented) and intermediate
reassembly is not viable, can routers still check content authenticity?
\end{minipage}}}\medskip

In other words, if verifying integrity/authenticity is the main reason for intermediate reassembly,
is there a way to obtain the former while avoiding the latter?

\subsection{Fragment Delivery Order}	
One important issue relevant to intermediate reassembly and to the
proposed technique (Section~\ref{securefrag}) is whether fragments are always 
delivered in transmission order between any two adjacent NDN routers.

Clearly, reassembly is easier if ordered delivery can be guaranteed. In a hypothetical
network setting where NDN is universally deployed directly on top of the physical network 
links, ordered delivery of content fragments might be a reasonable assumption.
\footnote{All content fragments traverse, in reverse, the very same path taken
by an interest.} However, certain connectivity and communication choices make ordered
delivery less likely. For instance, if adjacent routers support multiple/parallel physical links with variable
speeds, it is possible that an earlier-transmitted fragment is received later than
a later-transmitted one. Also, an error on one of the links might cause the same
situation even if link speeds are comparable. Even without multiple links, if 
pipelined data-link layer transmission is used, especially over the wireless channel, 
one fragment could be corrupted and discarded and the next one could be received
intact, resulting in the latter being received first.

\subsection{Incremental or Deferred Fragment Caching?}
Recall that one of the key features of NDN is router-based content caching. This is not,
strictly speaking, a hard requirement, however, it is expected that each NDN router will
maintain a Content Store (CS), i.e., a cache, of a certain size.

A router that employs intermediate reassembly can defer the decision to cache
content until it receives all fragments and, optionally, verifies overall content integrity and/or 
authenticity. Whereas, a router that employs {\em cut-through switching} of individual fragments
has a choice to either: (1) cache fragments incrementally as they arrive, or (2) defer caching (i.e., buffer
fragments) until all fragments arrive and, optionally, their overall content  integrity and/or authenticity is verified.
Assuming that most content is authentic, the former optimizes the common case of quickly caching the 
last fragment once optional security checks are performed. On the other hand, incremental caching
may complicate matters, since it might, depending on the specific cache architecture, 
involve non-contiguous caching of related fragments.

If deferred caching is used, another fragmentation-and caching-related issue is how to store 
fragments? One possibility is to store them in the same form they arrive. This might work
if no re-fragmentation is performed locally. Otherwise, it might make sense to store fragments 
in the form they are forwarded. This gets more complicated in case of collapsed interests, i.e., 
when content needs to be forwarded out on multiple interfaces with different MTUs. Another 
approach would be to proactively re-fragment cached fragments for all possible link MTUs on
the router. This pre-fragmentation would reduced delay at the cost
of additional buffer space. We believe this issue deserves
further consideration; which is beyond this paper's scope.

\section{Secure Fragmentation}\label{securefrag}
This section describes a scheme called FIGOA: Fragmentation with Integrity 
Guarantees and Optional Authentication. It supports arbitrary intermediate fragmentation \cite{mogul}
of content while preserving security, without requiring intermediate reassembly.\footnote{A variant
of FIGOA can be used in conjunction with intermediate reassembly, with the key advantage
of faster cryptographic processing.}
FIGOA allows free mixing of routers that do not perform intermediate reassembly with those that do. 
It is primarily geared for routers that support cut-through switching and maintain dedicated storage
for buffering content fragments, distinct from Content Store.
While cut-through fragment switching is generally beneficial, it complicates signature 
verification, as discussed in Section~\ref{reass}, FIGOA addresses this problem by using 
delayed authentication.

\subsection{Delayed Authentication}
Delayed authentication is an old method described in \cite{Tsudik89}. Its goal was to: 
{\em ``reconcile fragmentation and dynamic routing with network-level authentication in IP gateways.''}
The essence of delayed authentication is that a given packet's authenticity can be obtained from 
authenticity of its fragments. Packet authentication is computed incrementally as individual fragments 
are received (possibly out of order), processed and forwarded by a router. This requires a queue
for each partially received packet that maintains the current state of partial verification. For every 
fragment, incremental  verification is performed, queue state is updated and the fragment is forwarded. 
Upon receipt of the final fragment, the router completes verification. If it succeeds, the final 
fragment (called a ``hostage'') is forwarded. Otherwise, it is discarded along with the entire queue. 
The end-result is that the destination receives
the packet in its entirety only if it is verified by the router.\footnote{Recall that, in IPv4, the destination  
must flush all fragments of a packet that it can not reassemble, either due to a time-out or another error.}

Main differences between delayed authentication in its original IPv4 context \cite{Tsudik89} 
and our proposed use in NDN are as follows:
\begin{compactitem}
\item Symmetric routing: unlike IP, where fragments of the same IP packet might travel via different paths,
fragments of the same NDN content are guaranteed to follow the same sequence of NDN routers, retracing
PIT state set up by a preceding interest. This results in much higher probability of ordered fragment delivery 
and faster time-outs in cases of lost or corrupted fragments. Note that it is the responsibility of consumers to 
re-request the entire content in case of lost or corrupted fragments.
%
%
\item Not just ingress routers: delayed authentication was initially designed for ingress routers (i.e., border
routers of the destination's AS). In NDN, any intermediate router can unilaterally 
choose to perform delayed authentication.
\item Possible intermediate reassembly: in IP, only the destination reassembles fragments, whereas, any
intervening router can decide to reassemble whether or not it decides to do cut-through forwarding.
\item Signatures instead of CBC-based MAC: delayed authentication was initially proposed for authenticating
IP packet traffic flowing between two hosts (in two stub AS-s) that share a symmetric key. The same key
is shared with the ingress router. Actual packet authentication is attained via a message authentication code (MAC)
based on the chained-block cipher (CBC) mode of symmetric encryption. The main idea is to insert 
intermediate MAC values thus allowing incremental authentication of fragments. Whereas, in the NDN context,
MAC-s are not viable, since doing so would require sharing a symmetric key among all intervening routers. 
\end{compactitem}

\subsection{Hash Functions}
The last item above -- the use of signatures -- is what most distinguishes delayed authentication in NDN from its 
IP counterpart. NDN routers do not use symmetric cryptography for packet authentication. Even if they did, 
assuming a key shared among (possibly all) routers that forward a given content is unrealistic. 
The only means of authenticating content in routers is by verifying signatures.
This prompts the question: how to reconcile delayed authentication (of fragments) with signatures?

We approach this issue by observing that a signature is computed over a fixed-size 
hash digest (or simply hash) of content, i.e., using the so-called ``hash-and-sign'' paradigm.  
A hash provides integrity while a signature of a hash provides authenticity or origin authentication.
The underlying cryptographic hash function \(H(\cdot)\) must satisfy a set of standard 
properties \cite{menezes1996handbook}. Unlike a MAC or a keyed hash \cite{hmac}, a hash function requires 
no secret key and can be computed by anyone. 

Most modern hash functions operate on input of practically 
any\footnote{We consider $2^{64}$ or $2^{128}$ bits as ``practically any''.} 
size. They typically use an iterative model (also known as the Merkle-Damgard construction), whereby 
input is broken into a number of fixed-size blocks and is processed one block at a time by an internal
compression function \(HC(\cdot)\). The latter forms the core of the hash function; after processing each block, 
it produces an intermediate value (internal state that we call $IS$) that is usually of the same size as the 
final hash. In case of the first block,
the intermediate state is fixed and referred as the Initialization Vector (IV). The last 
block is typically padded with zeros followed by the total input size in bits. For example, 
the well-known SHA-256 \cite{sha} operates on 512-bit blocks, maintains 256-bit internal state and yields 
a 256-bit hash. 

As described below, in constructing FIGOA, we take advantage of internal state produced by the 
underlying compression function \(HC(\cdot)\). The main idea is to include, in each fragment, the internal
state of the hash function {\bf up to}, but not including, that fragment. This allows incremental 
hashing of each fragment without having received either preceding or subsequent fragment(s).

We assume that the absolute minimum MTU of any link or interface
that takes advantage of FIGOA is at least one block of data, one block of internal state and whatever
size is needed to accommodate a content fragment header (i.e., content name, flags, etc.). More precisely,
we assume that any fragment must carry at least a header, internal state and some data
represented as (at least one) some blocks of data. All data must be aligned with block
boundaries.

\subsection{FIGOA Description}\label{figoa}
\begin{table}[tbp]
\centering
\small
\resizebox{\linewidth}{!}{%
\begin{tabular}{|c|l|}
\hline
$\beta$  & block size of $HC(\cdot)$ \\
\hline
$CO^{n}$		& Raw (unsigned) content of total size $n$ bits. \\
\hline
$SIG(CO^n)$		& Producer's signature on $CO^n$. \\
\hline
$\overline{CO}^N$ & Signed version of $CO^n$ of size $N=n+|SIG(CO^n)|$ bits. \\
\hline
$b_{v,s}$		& Contiguous component of $\overline{CO}^N$ where $0\leq~v<N$, i.e., $b_{v,s}$ \\
				& represents $s$ bits, starting with offset $v$ and ending with offset \\
				& $v + s - 1$, inclusive. $s$ and $v$ are multiples of $\beta$. \\
\hline
$CF^N_{v,s}$	& Fragment of $\overline{CO}^N$ that carries $b_{v,s}$. \\
\hline
$IS_v$ & Internal state of $HC(\cdot)$ after processing $v$ bits of input. \\
       & $v$ is a multiple of $\beta$ \\
\hline
$hs$				& Fragment header size, includes: \\
				& content name, $v$, $s$ and $IS_v$. See Section~\ref{implementation} for details. \\
\hline
$o\mathcal{MTU}$		&  MTU of router's outgoing interface.\\ \hline
$ao\mathcal{MTU}$		&  $o\mathcal{MTU}$ adjusted for fragment header size $hs$,\\ 
                                          & i.e., $ao\mathcal{MTU}=o\mathcal{MTU}-hs$ \\
\hline
$\mathbb{F}$		& Set of content fragments. \\
\hline
$\mathbb{B}$		& Temporary buffer storing all fragments received so far. \\
\hline
\end{tabular}
}%
\caption{Notation}
\label{tab:notation}%
\vspace{-0.4cm}
\end{table}

From here on, we use additional notation reflected in Table~\ref{tab:notation}.
The proposed scheme includes three main tasks, described separately below.

\subsubsection{Content Fragmentation}
This task, shown in Algorithm~\ref{alg:fragment-content}, is triggered whenever an NDN node 
(router or producer) needs to forward a content object larger than $o\mathcal{MTU}$. 
Each resulting fragment $CF^N_{v,s}$ includes: (1) $s$ bits of original content -- $b_{v,s}$, 
(2) starting offset $v$, and (3) $IS_v$ -- intermediate state, i.e., output of $HC(\cdot)$ on inputs of: 
$IV$ and $b_{0,v-1}$.\footnote{In the very first fragment, $v=0$ and $IS_v=IV$.} ($IS_v = HC(IV, b_{0, v-1})$.)
To simplify presentation, Algorithm~\ref{alg:fragment-content} makes two assumptions:
{\bf First}, $ao\mathcal{MTU}$ is a multiple of $\beta$, i.e., $ao\mathcal{MTU}=s * \beta$.
{\bf Second,} $N$ (signed content size) is a multiple of $ao\mathcal{MTU}$, i.e., $N=k * ao\mathcal{MTU}$, 
which makes all fragments of equal size.

\begin{algorithm}[ht!]
\caption{Fragment-Content} \label{alg:fragment-content}
\begin{algorithmic}[1]
\scriptsize
\STATE {\bf Input:} signed content $\overline{CO}^N = b_{0,N-1}$, $ao\mathcal{MTU}$, $IV$, $HC(\cdot)$
\STATE {\bf Output:} $\mathbb{F}$
\STATE $\mathbb{F}:=\emptyset, \; v=0,\; IS_v=IV$
\STATE $s = ao\mathcal{MTU} / \beta, \; k= N / s $
\FOR   {$i=0, \; i < k, \; {i++} $}
	\STATE $CF^N_{v,s} := \left< \;  v, b_{v,s}, IS_v \right>$ 
	\STATE $\mathbb{F} := \mathbb{F} \cup CF^N_{v,s}$
	\STATE $IS_v := HC ( IS_v, b_{v,s} )$
	\STATE $v = v  + s$
\ENDFOR
\STATE Output $\mathbb{F}$
\end{algorithmic}
\end{algorithm}

\subsubsection{Fragment Re-fragmentation}
This task 
is very similar to the initial fragmentation task,
except that it is performed only by NDN routers, and on content fragments, instead of content objects.


\subsubsection{Content Verification}
As mentioned earlier, FIGOA provides integrity/authenticity for fragments received 
in any order.  Recall that a router or a consumer can unilaterally decide 
whether to either: (1) incrementally verify integrity of each fragment as it is received, or (2) defer 
overall verification until all fragments are received. Regardless of the choice,  a router 
should forward each fragment in a cut-through fashion, i.e., without waiting for others to arrive. 
Moreover, a node receiving fragments should store them in a buffer until the last fragment arrives
and (final or overall) verification is performed. (See Section~\ref{authentication}.) 

When a router performing incremental fragment verification receives $CF^N_{v,s}$,
one of the following cases occurs:
\begin{enumerate}
\item $CF^N_{v,s}$ is the very first received fragment. A new buffer $\mathbb{B}$ is created where $CF^N_{v,s}$ is placed. 
$IS^*_w=HC(IS_v, b_{v,s})$ is computed and stored.
\item Neither previous $CF^N_{u,s}$ (for $v = u + s$) nor next $CF^N_{w,s}$ (for $w = v + s$) fragment 
is in the buffer. $CF^N_{v,s}$ is placed in $\mathbb{B}$. $IS^*_w=HC(IS_v, b_{v,s})$ is computed and stored.
\item $CF^N_{u,s}$ is in the buffer (along with $IS^*_v$) but $CF^N_{w,s}$ is not. 
$IS^*_v$ must match $IS_v$ in $CF^N_{v,s}$. $IS^*_w=HC(IS_v, b_{v,s})$ is computed and stored.
\item $CF^N_{w,s}$ is in the buffer but $CF^N_{u,s}$ is not. 
$IS^*_w=HC(IS_v, b_{v,s})$ is computed and must match $IS_w$ from $CF^N_{w,s}$.
\item Both $CF^N_{u,s}$ and $CF^N_{w,s}$ have already been received.
$IS^*_v$ must match $IS_v$ in $CF^N_{v,s}$
$IS^*_w=HC(IS_v, b_{v,s})$ is computed and must match $IS_w$ from $CF^N_{w,s}$
\end{enumerate}
Once the last fragment is received, authenticity of the entire content can be finally verified. 
If verification fails, the last fragment is dropped, the PIT entry is flushed, and nothing is cached. 
The same applies for any failed check in the 5 cases above.
This process is illustrated in more detail in 
Algorithm~\ref{alg:verify-fragment}. We assume that routers perform incremental verification
of fragments and verify reassembled content signature. If signature verification is not possible,
routers must verify that the reassembled content hash matches the original content hash included in 
every fragment (see Section~\ref{implementation} for details.)

\begin{algorithm}[ht!]
\caption{Verify-Fragment} \label{alg:verify-fragment}
\begin{algorithmic}[1]
\scriptsize
\STATE {\bf Input:} received $CF^N_{v,s}$, associated PIT entry $e$, $HC(\cdot)$
\STATE {\bf Output:} no output
\IF {is\_first($CF^N_{v,s}$)}
	\STATE $\mathbb{B}:=$ get\_new\_buffer();
\ENDIF
\STATE INSERT $CF^N_{v,s}$ in $\mathbb{B}$
\STATE STORE $IS^*_w = HC(IS_v, b_{v,s})$
\IF {$CF^N_{u,s} \in \mathbb{B}$ \AND $IS^*_v \neq IS_v$ in $CF^N_{v,s}$}
	\STATE {\bf goto} {\em CleanUp}
\ENDIF
\IF {$CF^N_{w,s} \in \mathbb{B}$ \AND $IS^*_w \neq IS_w$ of $CF^N_{w,s}$}
	\STATE {\bf goto} {\em CleanUp}
\ENDIF
\IF {is\_not\_last($CF^N_{v,s}$)}
	\STATE FORWARD $CF^N_{v,s}$ according to $e$
\ENDIF
\IF {content\_complete()}
	\STATE $\overline{CO}^N:=$ assemble($\mathbb{B}$)
	\IF {verify\_sig($\overline{CO}^N$)}
		\STATE FORWARD $CF^N_{v,s}$ according to $e$
		\STATE CACHE $\overline{CO}^N$
		\RETURN
	\ENDIF
\ENDIF
\STATE {\em CleanUp:} FLUSH $\mathbb{B}$ and $e$
\end{algorithmic}
\end{algorithm}

\subsection{Examples}\label{figoa-example}
We now describe FIGOA via two operational examples:
{\bf In the first example,} consider a situation where $\overline{CO^N}$ 
has two fragments: $CF^N_{0,s}$ and $CF^N_{s,s}$, i.e. $N = 2 \times s$. A router $R$ first receives $CF^N_{0,s}$.
First, $R$ invokes \(HC(\cdot)\) iteratively and computes $IS^*_s = HC(IV, b_{0,s})$.
Then, it forwards $CF^n_{0,s}$ out on the interface(s) reflected in the corresponding PIT entry. 
$R$ creates a buffer for $\overline{CO}^N$ where it records the fact that it received the first $s$ bits of
content, along with the computed $IS^*_s$. Now, $R$ receives $CF^N_{s,s}$. It compares stored $IS^*_s$ with 
$IS_s$ carried in $CF^N_{s,s}$; if they do not match, $R$ discards the buffer and
flushes the corresponding PIT entry. Otherwise, it invokes
\(HC(\cdot)\) iteratively and computes $IS^*_N = HC(IS_s, b_{s,s})$. At the end, $R$ extracts $SIG(CO^n)$ (from the received content), 
and computes a putative hash $H'$ of entire reassembled $CO^{n}$.
Finally, $R$ verifies whether $SIG(CO^n)$ is the producer's signature on $H'$. If so,
$CF^N_{s,s}$ is forwarded; otherwise, it is discarded along with the buffer and the PIT
entry. 
A similar process takes place if $CF^N_{0,s}$ and $CF^N_{s,s}$ arrive out of order.
$R$ first receives $CF^N_{s,s}$. Using $IS_s$ carried in this fragment, $R$ invokes
\(HC(\cdot)\) iteratively on each block of data and terminates with $IS^*_N$.
Next, $R$ forwards $CF^N_{s,s}$. Then, $R$ creates a buffer for $\overline{CO}^N$ where it 
records the fact that it received the last $N-s$ bits (which is in fact the last $s$ bits) of
content, along with $IS_s$ and $IS^*_N$. Now, $R$ receives 
$CF^N_{0,s}$. It invokes 
\(HC(\cdot)\) iteratively and computes $IS^*_s = HC(IV, b_{0,s})$ which should match $IS_s$ received earlier 
as part of $CF^N_{s,s}$: if they do not match, $R$ discards the buffer
and the PIT entry. Otherwise, $R$ computes a putative hash $H'$ of entire reassembled $CO^{n}$, 
extracts $SIG(CO^n)$ and verifies whether it is the producer's 
signature on $H'$. If so, $CF^N_{0,s}$ is forwarded; otherwise, it is discarded along with the 
buffer and the PIT entry. 
{\bf The second operational example} involves $R$ receiving a fragment $CF^N_{x,s}$ of  
content $\overline{CO}^N$. The total size of this fragment is $(s + hs)$ bits. 
Suppose that, after processing this fragment as in the first 
example, $R$ needs to forward it out on an interface with $o\mathcal{MTU}$ smaller than the total size
of $CF^N_{x,s}$, e.g., $R$ needs to re-fragment it into two sub-fragments.
$R$ creates $CF^N_{x,s'}$ with $IS_x$ and $CF^N_{y,s'}$ with $IS_y$; where (1) $s' < s$, 
(2) $y = x + s'$, (3) $IS_x$ is simply copied from $CF^N_{x,s}$, and (4) $IS_y = HC(IS_x, b_{x, s'})$.
This example aims to show that $R$ can easily re-fragment already-fragmented content while
preserving overall content integrity.

\begin{figure}
\centering
\includegraphics[width=\columnwidth]{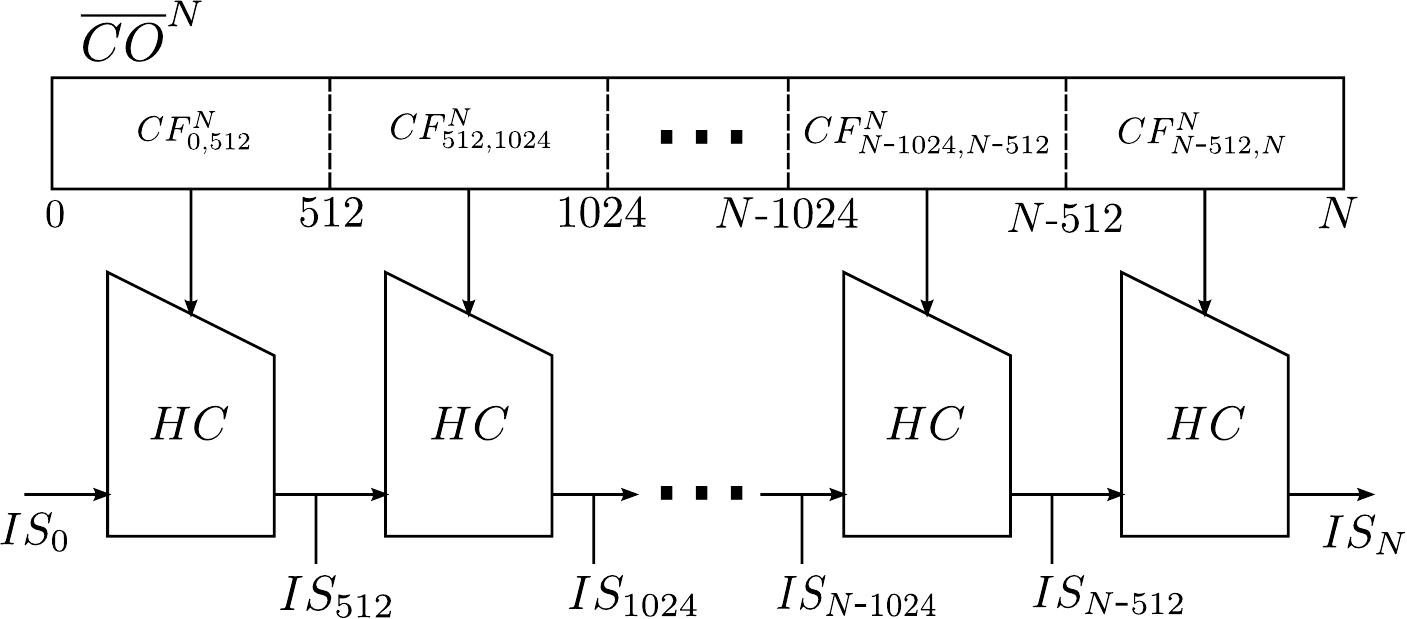}
\caption{Implementing Merkle-Damgard Construction to Generate Content Fragments}
\label{fig:scheme}
\end{figure}

Figure~\ref{fig:scheme} demonstrates how to use any hash function based on the Merkle-Damgard construction 
to generate content fragments. The hash function used in Figure~\ref{fig:scheme} is SHA-256, and the length of 
input is discarded at the end of construction to simplify demonstration.

\subsection{Content Authentication}\label{authentication}
Although trust and key management are out of the scope of this paper,
we can not ignore the fact that authenticating a content object requires not only the
presence of a signature, but also availability of a public key which must somehow be
trusted~\cite{ghali2014network}. Recall that NDN stipulates that public keys are encapsulated in 
named and signed content objects, i.e., a form of a certificate. Also, NDN allows the public key to be either: 
(1) referred to by name within a content object header, or (2) enclosed with the content object itself, using the 
\texttt{KeyLocator} field.\footnote{However, no trust management architecture 
is defined in NDN.} In the former case, unless the referred public key is already cached, the
router presumably must fetch it by name, i.e., issue an interest for it. 
This is a burdensome task that routers should not perform, for obvious reasons.

\subsection{Security Analysis}\label{figoasec}

Security of FIGOA is based on that of delayed authentication (DA).

We say that $H(\cdot)$ is 
constructed using the Merkle-Damgard construction using its inner compression function $HC(\cdot)$ 
as a building block. If $HC(\cdot)$ is collision-resistant, then so is 
$H(\cdot)$.

A function $F$ is collision-resistant if it is 
``computationally infeasible'' to find inputs $x \neq y$ such that $F(x)=F(y)$. See \cite{menezes1996handbook} for 
information regarding Merkle-Damgard construction and hash-and-sign paradigm; 
also known as ``digital signature with appendix."

\ignore{\footnote{A function $F$ is collision-resistant if it is 
``computationally infeasible'' to find inputs $x \neq y$ such that $F(x)=F(y)$. See \cite{hac} for 
information regarding Merkle-Damgard construction and hash-and-sign paradigm; 
also known as ``digital signature with appendix."}}

A signature computed via hash-and-sign over an unfragmented content object 
is considered secure. Whereas, with DA, a content object is fragmented and 
we arrive at the final hash of the content packet by incrementally hashing its fragments.

To subvert DA we consider an adversary who is given a valid $\overline{CO}^{N}$ with signature 
$SIG(CO^n)$. The goal of the adversary is to send to some router $R$ a sequence of fragments, 
$CF'^{N'}_{x_0 = 0,s},CF'^{N'}_{x_1,s},\ldots,CF'^{N'}_{x_{k},s}$ ($x_{i + 1} = x_i + s, 0 \leq i \leq k - 1$) 
corresponding to $\overline{CO'}^{N'} \neq \overline{CO}^{N}$ with $H({CO'}^{x_{k}}) = H(CO^{n})$. 
Recall that $CF^N_{v,s}$ embodies intermediate state $IS_{v}$ of the hash function computed up to, 
but not including, $v$ bits of the entire content. 

First, consider fragments arriving in order. $R$ receives $CF'^{N'}_{0,s}$, initializes $HC$ with $IV$, and computes 
and retains ${IS'^*}_{x_1} = HC(IV, b_{0,s})$. Now, when 
$R$ receives subsequent fragments $CF'^{N'}_{x_i, s}, i = 1, \ldots, k$, it compares the current (computed)
${IS'^*}_{x_1}$ with ${IS'}_{x_1}$ contained in $CF'^{N'}_{x_1, s}$. If they match, $R$ invokes 
$HC(\cdot)$ iteratively over each block of $CF'^{N'}_{x_i,s}$ using ${IS'}_{x_i}$ as the starting intermediate state, 
and compares obtained ${IS'^*}_{x_{i+1}}$ with ${IS'}_{x_{i+1}}$ in $CF'^{N'}_{x_{i+1},s}$.
This process is exactly the same as computing $H(\cdot)$ over entire $\overline{CO'}^{N'}$. 
If $\overline{CO'}^{N'} \neq \overline{CO}^{N}$, the adversary must have found a collision 
for $H(\cdot)$, which violates our collision-resistance assumption.

Now, assume that fragments arrive out-of-order. $R$ receives $CF'^{N'}_{x_i,s}$. It can readily compute 
${IS'^*}_{x_{i+1}}$ by invoking $HC(\cdot)$ over the blocks starting with ${IS'}_{x_i}$. 
$R$ retains ${IS'}_{x_i}$ as part of its state until ${IS'^*}_{x_i}$ is computed 
(using all previous fragments) and matched. If ${IS'^*}_{x_i}$ 
has been already computed, then $R$ must have invoked $HC(\cdot)$ over the data in 
$CF'^{N'}_{x_{i-1},s}$ using ${IS'}_{x_{i-1}}$. If $R$ arrives at the final hash 
output and its state contains {\em only} ${IS'^*}_{N'}$, then $R$ has compared ${IS'^*}_{x_i}$ with 
${IS'}_{x_i}$ for $i = 1, \ldots, k$ such that each match was successful. In other words, 
all fragments have arrived and  $i = 1, \ldots, k, {IS'^*}_{x_i} = {IS'}_{x_i}$. We observe that 
the set of equations that must be \emph{satisfied} here is exactly the same as that in the in-order-arrival 
case. Therefore, the same argument applies.

\section{Implementation}\label{implementation}

In this section, we describe the implementation of FIGOA in CCNx version 0.8.2 \cite{ccnx} (latest version while writing this paper.) Our implementation performs fragmentation with cut-through switching, and intermediate reassembly. We strive to remain as consistent and compatible with the existing CCNx codebase, without changing the architecture or design except to support fragmentation. Due to lack of support for signature verification and key management in the implementation of the CCNx codebase, our implementation does not support signature verification of content objects processed by routers. However, it can naturally be extend to authentication should this feature becomes present in a future CCNx version.

CCNx is an open source content-centric networking stack developed by Palo Alto Research Center (PARC). The software suite comprises of a forwarder ({\em ccnd}) and client ({\em libccn}) implemented in the C programming language. A client for Java is also available. We refer to \cite{ccnx-protocol} for more detailed specifications regarding the CCNx protocol.

Our implementation only requires modification of the forwarder code. Our design limits fragmentation, reassembly, and cut-through switching for outgoing interfaces. Therefore, a forwarder must reassemble fragments prior to forwarding over the content to the application.

To implement fragmentation, we introduce a new type of NDN packet, \ContentFragment. This packet is used for both initial fragmentation of content objects and re-fragmentation of content fragments. The structure of \ContentFragment\ contains the following fields:
\begin{compactitem}
\item {{\Name}:} identical to content name without an additional implicit component digest.
\item {{\ContentObjectSize}:} size of the original content object before fragmentation takes place.
\item {{\InternalState}:} stores internal state of a SHA-256 computation up to {\PayloadOffset} of the content.
\item {{\PayloadOffset}:} specifies where the fragmented data begins with respect to the unsigned content.
\item {{\PayloadSize}:} size of fragment payload, which is a multiple of 512-bits (the input block size for SHA-256) except for the last fragment.
\item {{\ContentDigest}:} contains the digest of the original content object. Appending this digest to the end of {\Name}, forms the content's unique name. This field allows router to match fragments with interests (in PIT) containing the content digest as part of their names. Moreover, for routers not verifying content signature, this field must match the hash computed after reassembling the content.
\item {{\Payload}:} fragmented data of the content.
\end{compactitem}
%

Once all fragments are received and the content is reassembled, the router caches it, if its integrity is verified.

The above format lends itself to natural re-fragmentation. If a \ContentFragment\ requires further fragmentation only \InternalState, \PayloadOffset, \PayloadSize, 
and \Payload\ fields needs to be adjusted, reflecting the new fragments. This prevents nested fragments and simplifies reassembly. Thus, increasing routers performance and reducing consumers end-to-end latency.

To evaluate our implementation, we compare its performance to an unmodified version of CCNx 0.8.2. This version (similarly to the current NDN testbed) is running as an overlay network on top of TCP or UDP. When TCP is used to connect CCNx nodes, content larger than the negotiated MTU (at the connection setup) will be segmented by TCP. This reduces the chance of IP fragmentation to take place unless the MTU dropped to a smaller value at an intermediate router. On the other hand, when UDP is used, IP will be responsible of fragmenting and reassembling content. In this case, every CCNx node receives the whole content object from the UDP socket after reassembly is performed by IP. For the purpose of our experiments, we use UDP as a transport layer protocol to compare the performance of our FIGOA implementation to that of IP fragmentation.

\section{Evaluation}\label{evaluation}
We employ a server equipped with 8-core Intel i7-3770 CPU at 3.40GHz and 16GB of memory. The server runs Ubuntu 12.10 and KVM hypervisor to run virtual machines. We construct a testbed by provisioning virtual machines to act as CCNx nodes interconnected in the same LAN and NATed by the host server. Each node is connected to virtual Ethernet interface at 100Mbps and MTU set to 1500 bytes.

Experiments are run on a 3, 4, and 5 nodes linear-topology. The first hop acts as consumer sending interests with a specific content published by the last hop (the producer). For each topology, we run the experiments in which consumers request content with data size of 1, 2, 4, 8, 16, and 32 KB. The reason we chose a linear-topology is because content objects and fragments always follow the same path, in reverse, of preceding interests.

\begin{figure*}
\subfigure[3 nodes]
{
	\includegraphics[width=0.65\columnwidth]{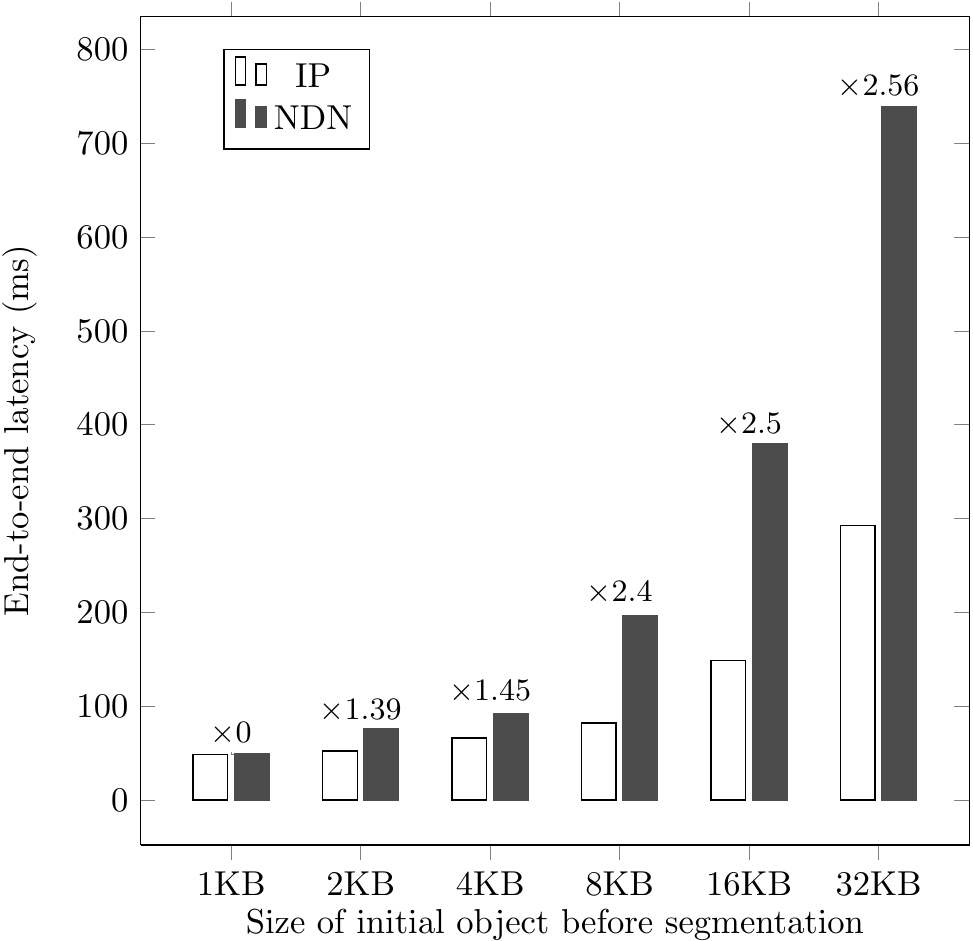}
	\label{fig:eval-3nodes}
}
\hspace{-0.1cm}
\subfigure[4 nodes]
{
	\includegraphics[width=0.65\columnwidth]{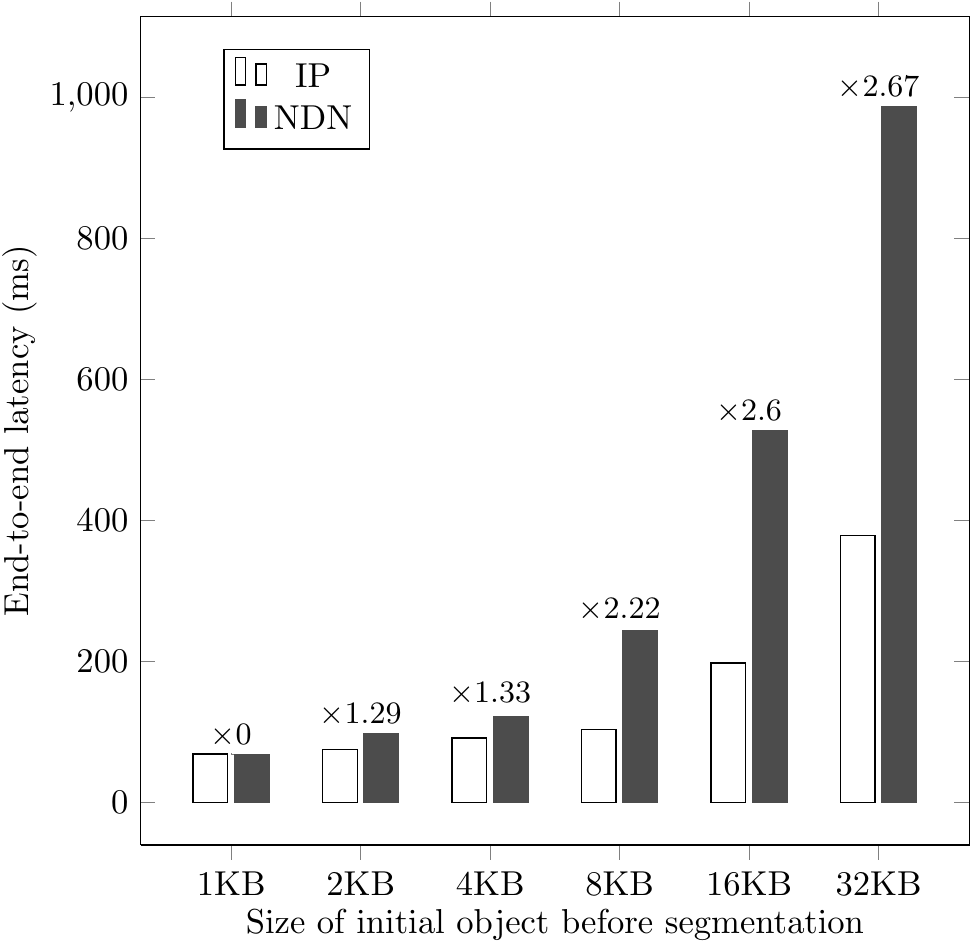}
	\label{fig:eval-4nodes}
}
\hspace{0.15cm}
\subfigure[5 nodes]
{
	\hspace{-0.4cm}
	\includegraphics[width=0.65\columnwidth]{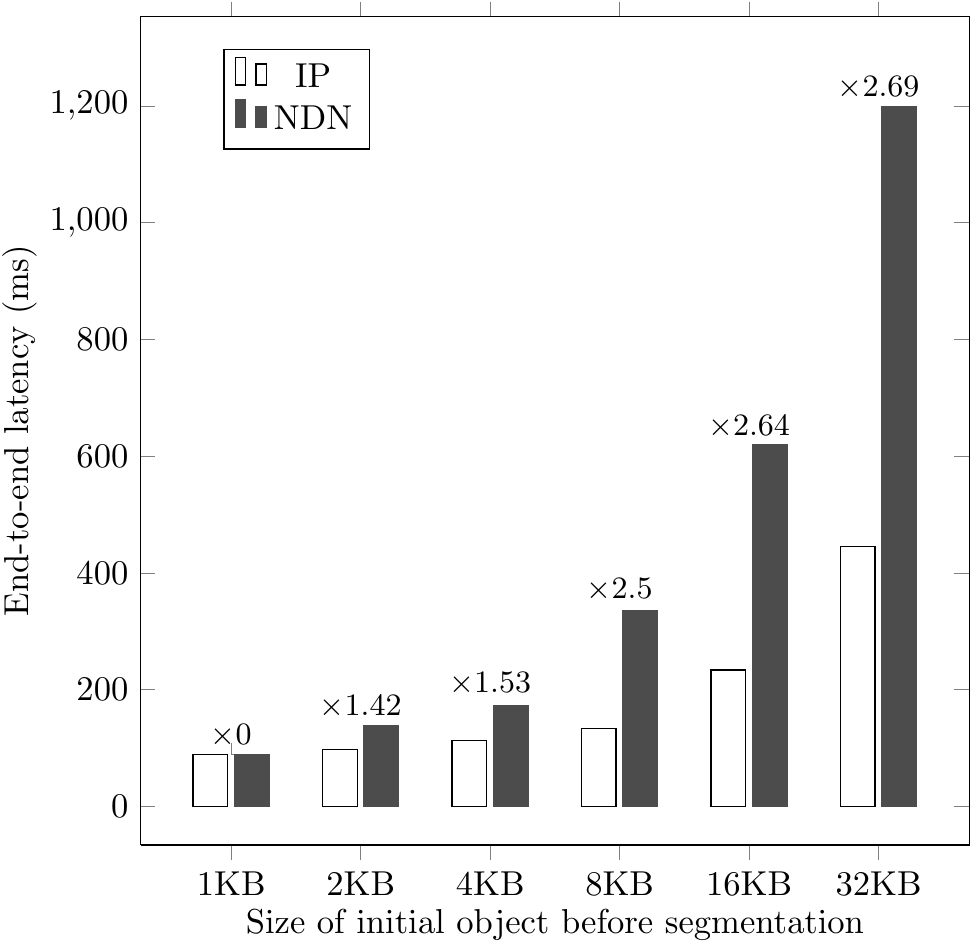}
	\label{fig:eval-5nodes}
}
\caption{End-to-end latency of various sized content retrieval. IP represents the unmodified version of CCNx and NDN represents FIGOA. The values above the bars represent the difference between NDN and IP fragmentation (NDN / IP).}
\label{fig:evaluation-results}
\end{figure*}

Results are shown in Figure~\ref{fig:evaluation-results} demonstrating the average consumer end-to-end latency measured from many repeated experiments.\footnote{All nodes start with an empty cache at the beginning of every experiments.} For all settings, IP performs consistently better than our cut-through approach. The bottleneck of FIGOA is that routers need to perform additional processing to compute the hash of every fragment. Since all computations are currently performed in software, these results make sense. However, we believe that once NDN/CCN is deployed as a replacement of IP, all nodes (especially routers) will be capable of performing hash computation at the hardware level at a rate much faster than what is shown in Figure~\ref{fig:evaluation-results}. 

\begin{figure}[ht]
\centering
\includegraphics[width=0.65\columnwidth]{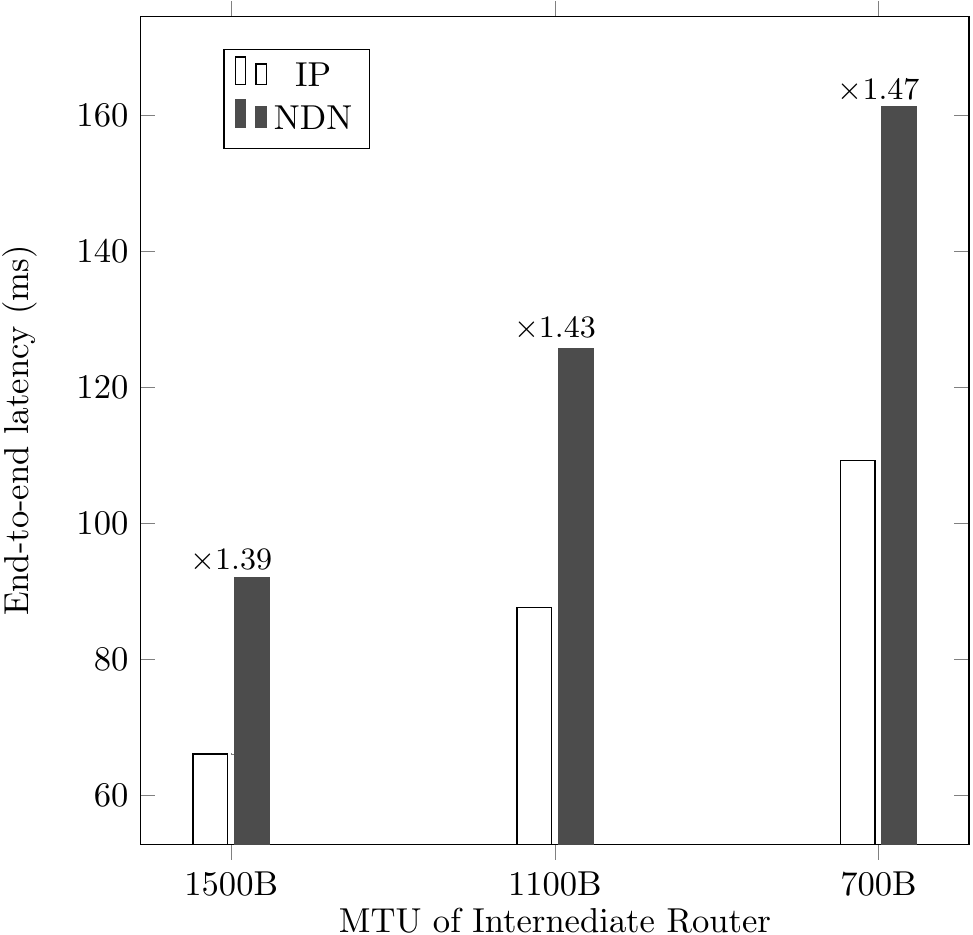}
\caption{End-to-end latency of various MTU-s at intermediate routers for content size 4KB. 
IP represents the unmodified version of CCNx and NDN represents FIGOA. 
Values above the bars represent the difference between NDN and IP fragmentation (NDN/IP).}
\label{fig:3_nodes_refrag}
\end{figure}

We run another 3 nodes experiment that involves refragmenting fragments. We measure the end-to-end latency at the consumer for different values of intermediate router's MTU (1500, 1100, and 700 bytes). Consumer pulls content of size 4KB. In the case of MTU value equal to 1500 bytes, the content is fragmented (at the producer) into 4 fragments, each, except the last one, is of size 1152 bytes\footnote{Multiple of SHA-256 block size which is 64 bytes.} of effective fragment payload plus fragment header length. However, when MTU drops to 1100 bytes (at the intermediate router), the payload length of each outgoing becomes 768 bytes, leading to re-fragmentation of each fragment into 2 smaller ones. Similarly, each fragment is re-fragmented into 3 smaller fragments when MTU value drops to 700 bytes.

Results are shown in Figure~\ref{fig:3_nodes_refrag}. We can notice that when MTU value decreases, the end-to-end latency increases for a fixed content size. This is a logical conclusion due to the fact that smaller MTU leads to more processing imposed by re-fragmentation. Although re-fragmentation using FIGOA requires additional hash computations at each hop after where re-fragmentation occurs\footnote{Recall that IP re-fragmentation does not required hash computation.}, the ratio of NDN end-to-end latency to IP end-to-end latency is not increasing dramatically. The reason is due to the fact that when fragmentation happens at the IP layer, reassembling the content is required at every hop before it is delivered by the UDP socket to NDN. Since this is not the case when FIGOA is implemented, IP reassembly adds more end-to-end latency that compensates the additional hash computation overhead imposed by FIGOA.\footnote{Refer to Section~\ref{reass} for more details about the delay imposed by IP reassembly at each hope.}

\section{Related Work}\label{related}
Related work falls into several categories discussed here.
\subsection{Secure Fragmentation}\label{related-secure-fragmentation}
The first attempt to address security in IP fragmentation is  \cite{Tsudik89} which tackled a 
specific problem of how to authenticate, in egress/ingress routers, fragmented IP packets. 
A source host is assumed to  share a key with appropriate router(s). Two techniques are proposed: 
The first one is delayed authentication (DA) where an authenticating router verifies a packet MAC 
incrementally from its fragments. Since fragments of the same packet might flow through different 
routers, to prevent reassembly of a corrupted packet at the destination, an authenticating 
router holds one small fragment ``hostage" until authenticity of the entire packet authenticity is confirmed. 
The second scheme is an MTU probe mechanism that a source host can use to pre-segment a 
large packet into smaller authentic packets sized to the smallest MTU on the (current) path. 
Some extensions to \cite{Tsudik89} were later proposed in \cite{Popp93}: extended delayed authentication 
(EDA) requires fragments to always traverse the same path. 
\cite{Popp93} also provides a detailed comparison of several secure fragmentation techniques.

\cite{Partridge05} presents a secure fragmentation scheme for Delay-Tolerant Networks (DTN) \cite{rfc4838}. 
This scheme is referred to as ``toilet-paper" approach to securing fragments. The basic idea is that, 
prior to bundling, data is checkpointed into fragments using cryptographic hash at specified intervals. 
Hashes are included in a bundle and authenticated with a signature. Given a fragment, the hash, and the signature a gateway 
can authorize if the fragmented data should be delivered over the link. A variation of the scheme allows for variable increments 
of authentic fragments, allowing routers greater flexibility to choose fragment size potentially saving valuable link resources. 

An enhancement to the ``toilet-paper'' approach is presented in~\cite{asokan2007towards}. After bundle fragmentation occurs, senders build a hash tree and only sign the root node. Fragment verification requires the knowledge of this signature and $\log(n)$ hashes (where $n$ is the number of fragments). To verify the authenticity of all fragments, verifiers compute $n\log(n)$ hashes and a single signature verification, instead of $n$ hashes and $n$ signature verification operations. However, these approaches are not applicable in NDN since they both lack the support of in-network fragmentation. Moreover, in FIGOA, verifying all fragments requires the computation of only $n$ hashes and one signature verification.

Today, most networks employ IPSec \cite{rfc6071} to provide network-level authentication in IP networks. IPSec is compatible with both IPv4 and IPv6. IPSec operates in transport and tunnel mode. Transport mode is used by two hosts which establish a security association (exchanging keys) to authenticate and encrypt IP payloads. Tunnel mode allows the creation of secure Virtual Private Networks (VPN) which comprise of IPSec-enabled gateways that share bilateral security associations. Gateways secure and authenticate whole IP datagrams, encapsulating them as payloads for IP datagrams destined for processing between IPsec-enabled gateways.

Regardless of mode chosen, fragmentation of packets between IPSec-enabled hosts (gateways) occurs at the IP layer. Since IPSec authenticated/encrypted packets have a destination address of another IPSec-capable host (or gateway), it must undergo packet-level scrutiny which requires reassembly of the packet. In essence hop-by-hop reassembly at IPSec adjacent hosts ensures that security is not subverted.

\subsection{Fragmentation in ICN}\label{related-fragmentation-icn}
CCNx currently serves as a reference implementation for NDN. It currently supports TCP/UDP tunnels to interconnect forwarders. Fragmentation is relegated to IP, limiting the maximum packet size to that of IP. Hop-by-hop reassembly allows routers to
authenticate content (although not presently supported in the forwarder implementation), but at the increased cost of reassembly.

NDN Link Protocol (NDNLP) \cite{NDNLP} attempts to amend this issue while allowing operation over both link-layer and virtual transports, 
such as Ethernet and TCP/UDP. Fragmentation occurs for both interest and content packets.
It specifically features intermediate reassembly and therefore remains compatible with NDN security requirement, however
uses an incompatible packet format to support cut-through fragmentation which can result in incurred delay. NDNLP also
supports reliability layer.

The CCN-lite project \cite{ccn-lite} aims to provide a ``level-0" forwarder for CCN. It is compatible with the CCNx protocol
and provides a rudimentary implementation of the forwarder with simple data structures for PIT, FIB, and CS.
Native fragmentation and reassembly is supported over Ethernet and TCP/UDP. Fragments are identified by sequence number without any addressing scheme on per-fragment basis implying cut-through fragmentation is not supported. The fragmentation scheme also provides optional support for reliable fragments transmission.

CONET \cite{conet} is a derivative ICN of CCNx. In \cite{conet-transport-issues} a transport scheme called ICN 
Transport Protocol (ICTP) is specified, which implements TCP native to ICN. Similar to TCP, ICTP 
segments data to avoid further fragmentation. In essence, this provides cut-through delivery of fragments. 
Akin to TCP, this doesn't prevent fragmentation from occurring at a lower-layer. Unlike our scheme, ICTP does not
address content authentication at intermediate routers.

The NetInf project \cite{NetInf} is an emerging ICN architecture which supports location-independent named data objects (NDO) (similar to content objects in NDN/CCN). NDOs are signed and cacheable units. The project does not envisage a scheme for segmentation and relies on a ``convergence layer" (CL) to synthesize necessary services for heterogeneous transports used to connect NetInf gateways. The CL is delegated the responsibility of fragmentation and reassembly of NDOs. With no native fragmentation and reassembly scheme available, NetInf appears to rely on intra-hop reassembly for verification of NDO authenticity.

\section{Conclusion}\label{end}
Secure fragmentation is an important issue in NDN. It is 
complicated by the rule that each content object must be 
signed by its producer. Thus far, fragmentation of content objects has been 
considered incompatible with NDN since it precludes authentication of individual 
fragments by routers. 
In this paper, we showed that secure and efficient content fragmentation is both possible and 
advantageous in  NDN and similar architectures that involve signed content. We demonstrated a 
concrete technique (FIGOA) that facilitates efficient and secure content fragmentation in NDN,  
discussed its security features and assessed its performance. Finally, we described a prototype 
implementation and presented preliminary results.
\balance


\bibliographystyle{abbrv}
\bibliography{frag}

\end{document}